\documentclass{article}
\usepackage[labelfont=bf]{caption}
\usepackage[font={small}]{caption}
\usepackage[inkscapelatex=false]{svg}
\usepackage[]{pdfpages}

\usepackage{graphicx}%
\usepackage{dcolumn}%
\usepackage{bm}%
\usepackage[pdftex,colorlinks=true,linkcolor=blue,citecolor=blue]{hyperref}
\usepackage[utf8]{inputenc}
\usepackage{lipsum}  
\usepackage{amsmath}
\usepackage{authblk}
\usepackage{amsfonts}
\usepackage{placeins}
\usepackage[numbers, sort&compress]{natbib}  %
\usepackage{bibunits}
\usepackage{amssymb}
\usepackage{pdflscape} %
\usepackage{multirow}
\usepackage{mathrsfs}

    \usepackage[table, usenames, dvipsnames]{xcolor}
    \usepackage{makecell}
    \usepackage{boldline}
    \setcellgapes{3pt}

\usepackage{textgreek}      %

\usepackage[strict]{changepage} %
\usepackage{nicefrac}

\defaultbibliographystyle{naturemag}
\defaultbibliography{references}

\usepackage{lineno}
\usepackage{titling}

\begin{document}
\begin{bibunit}

\title{Stimulated Electro-optic Scattering}

\author{
    Violet Workman$^1$ and Gaurav Bahl$^2$\\
    $^1$ Department of Physics, \\
    $^2$ Department of Mechanical Science $\&$ Engineering, \\
    University of Illinois at Urbana–Champaign, Urbana, IL 61801 USA \\
}

\date{}

\maketitle

\begin{abstract}

Stimulated Brillouin Scattering (SBS) couples light to acoustic waves and underpins applications ranging from sensing and signal processing to quantum photonics. In solids, this interaction is generally attributed to photoelasticity and the motion of dielectric boundaries. Here we show that piezoelectricity opens an additional, previously unrecognized pathway that can substantially reshape Brillouin gain. This contribution arises from the combined action of linear electro-optic ``Pockels'' scattering and its reciprocal $\chi^{(2)}$-mediated optical drive, which couple optical fields to the electric fields that accompany mechanical motion in piezoelectrics. We formally incorporate these effects into the modern theory of SBS and apply it to monolithic lithium niobate waveguides. In addition to large enhancement and suppression of the Brillouin gain, we surprisingly find occurrences of large gain where conventional theory would otherwise have predicted zero Brillouin interaction. These instances are the first identification of purely electro-optic SBS, a process we term stimulated electro-optic scattering, as it only involves interactions between electromagnetic fields even though a mechanical excitation is present. Our results establish the electro-optic tensor as a key governing parameter for Brillouin interactions in piezoelectric media and provide a new degree of freedom for engineering photon-phonon coupling independent of photoelasticity and index contrast.

\end{abstract}

\section{Introduction}

Stimulated Brillouin Scattering (SBS) is a third-order nonlinear optical process arising from coherent interactions between optical photons and mechanical phonons \cite{chiao_stimulated_1964,boyd_nonlinear_2008}. SBS results from two simultaneous processes referred to as the optomechanical forward and back action \cite{wolff_stimulated_2015}. In the forward action, a mechanical excitation scatters light between optical modes. In the back action, scattered light beats together with the incident light, producing a force distribution that is phase-matched to either amplify or attenuate the mechanical excitation. SBS has found extensive use in applications such as narrow linewidth lasers \cite{chiao_stimulated_1964,li_characterization_2012} and linewidth sharpening \cite{debut_linewidth_2000,debut_experimental_2001}, optical phase conjugation \cite{zeldovich_connection_1995}, generating slow and fast light \cite{okawachi_tunable_2005,song_gain-assisted_2005}, Brillouin cooling \cite{bahl_observation_2012,kim_role_2017,chen_brillouin_2016} and excitation of phonon modes \cite{zhu_stored_2007,shin_control_2015,merklein_chip-integrated_2017}, producing nonreciprocal devices \cite{hwang_all-fiber-optic_1997,kim_complete_2017,kim_non-reciprocal_2015}, and optical wavelength conversion \cite{dong_optomechanical_2012,dong_optical_2015}.

Early analytical formulations of SBS primarily focused on plane wave scenarios, both for the photons and phonons, which is a good approach for bulk crystals \cite{chiao_stimulated_1964,boyd_nonlinear_2008}. In this picture, SBS can be treated as an intrinsic material property resulting from the combination of photoelastic scattering (forward action) and electrostriction (back action). This treatment was subsequently extended to consider other forward- and back-action pairs which occur in bulk crystals, such as roto-optic scattering (forward action) and electroconvolution (back action) \cite{nelson_new_1970}. A more thorough treatment considers the geometry-dependent guided optical and mechanical modes of bounded systems \cite{rowell_brillouin_1978,loudon_theory_1978,thomas_normal_1979}, e.g. optical fiber, thin films, waveguides, and whispering-gallery resonators. In strongly confined systems, it becomes essential to employ a full modal analysis as the Brillouin gain becomes strongly dependent on geometry \cite{kang_tightly_2009}. It was later demonstrated that nanoscale waveguides exhibit an entirely new pair of couplings contributing to SBS through the moving boundary effect (forward action) and radiation pressure (back action), which can dramatically enhance \cite{rakich_giant_2012} or diminish \cite{florez_brillouin_2016} the total gain. This can be generalized to cases where the involved optical and mechanical fields are tightly confined near material boundaries \cite{dostart_giant_2015}. These studies show the importance of two main considerations in the state of the art SBS theory. First, we must properly identify the optical and mechanical modes of the structure of interest. Second, we must account for the contribution of all forward- and back-action processes.

It has been known for some time that piezoelectric materials can host a unique acousto-optic forward-action due to the co-occurrence of the piezoelectric and electro-optic effects. This action was first taken into account for plane wave scenarios with the indirect photoelastic effect \cite{nelson_theory_1971}, which is a cascaded process of the piezoelectric coupling from acoustic to RF fields and the resulting electro-optic scattering from the generated fields. To include this action in a modern modal analysis, Shao et. al. \cite{shao_microwave--optical_2019} considered interactions between optical and hybrid \emph{electro-mechanical} modes, and treated the electro-optic effect as a single stage process similar to other bulk acousto-optic scattering effects. Our key insight is that linear electro-optic scattering is reversible through the $\chi^2$ nonlinearity, implying that optical modes can act back on electro-mechanical modes through their electric fields. Thus, together, the linear electro-optic and $\chi^2$ effects should act as a new forward- and back-action pair, a pair that that has not been incorporated into the modern guided mode theory of SBS.

In this work, we demonstrate that electro-optic scattering combined with the $\chi^{(2)}$ back action drastically modifies the Brillouin gain, suggesting a powerful new pathway for gain enhancement and suppression. We achieve this by developing a new full modal formulation of the Brillouin gain in piezoelectric materials that formally includes the effect of this new forward and back action pair, which we refer to as the electro-optic contribution to SBS. This marks a significant deviation from the state of the art SBS theory, which considers only interactions between optical and mechanical displacement fields \cite{wiederhecker_brillouin_2019, wolff_brillouin_2021}. To explore the role of the EO contribution, we perform fully anisotropic simulations including the piezoelectric coupling, and as familiar test cases, simulate the Brillouin gain for lithium niobate waveguides. We confirm that the EO contribution can significantly enhance or diminish the Brillouin gain depending on the choices of crystalline orientation, geometry, and optical polarization. More strikingly, we identify cases where the EO and $\chi^{(2)}$ effects are the only contributor to Brillouin scattering. We refer to this novel phenomenon as stimulated electro-optic scattering (SES), since although a mechanical displacement must be present, the forward action and back action involve only couplings between electromagnetic fields.

\section{SBS in piezoelectric waveguides}

\begin{figure}[t!]     
    \begin{adjustwidth*}{-0.5in}{-0.5in}
    \hsize=\linewidth
    \centering
    \includegraphics[page=1,clip, trim=0.6cm 18.5cm 2cm 3cm, width=\linewidth]{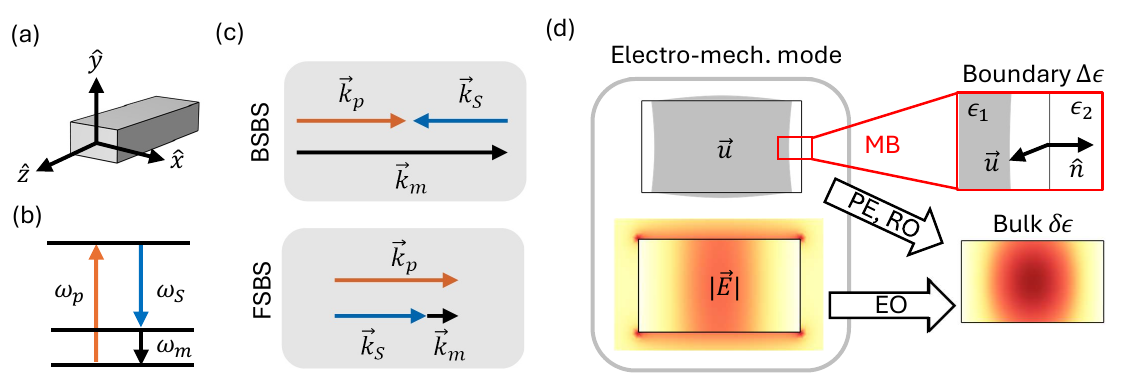}
    \caption{\textbf{SBS in piezoelectric waveguides. (a)} We examine waveguides with all waves propagating in the $\hat{z}$ direction. \textbf{(b)} SBS follows energy conservation, requiring $\omega_p=\omega_s+\omega_m$. \textbf{(c)} SBS also conserves momentum, resulting in two possible configurations. In backward SBS (BSBS), counterpropagating pump and Stokes beams are coupled by traveling wave phonons. In forward SBS (FSBS), copropagating pump and Stokes beams are coupled by standing wave phonons. \textbf{(d)} SBS in piezoelectric materials results from coupling between optical modes and hybrid electromechanical modes characterized by both displacement $(\vec{u})$ and electric $(\vec{E}^{mech})$ fields. The displacement field changes the permittivity at material interfaces through the moving boundary (MB) effect, and in the bulk through the photoelastic (PE) and roto-optic (RO) effects. The electric field also changes the bulk permittivity through the linear electro-optic (EO) effect.}
    \label{Calculation}
    \end{adjustwidth*}
\end{figure}

In this section, we develop a theory of SBS in piezoelectric waveguides accounting for the hybrid electromechanical nature of the acoustic wave and including the electro-optic contribution. We consider collinear propagation of two optical waves and one electromechanical wave in a waveguide oriented along the $\hat{z}$ direction (Fig.~\ref{Calculation}a). Following convention, we denote the higher frequency optical wave as pump, and the lower frequency optical wave as Stokes, and refer to the pump, Stokes and electromechanical waves by the subscripts $p$, $S$, and $m$ throughout. SBS satisfies the energy conservation condition $\omega_p=\omega_S+\omega_m$ (Fig.~\ref{Calculation}b). SBS also simultaneously conserves momentum $(\vec{k}_p=\vec{k}_S+\vec{k}_m)$, resulting in two well known cases illustrated in Fig.~\ref{Calculation}c. In backward SBS (BSBS), counterpropagating pump and Stokes photons are coupled by GHz frequency traveling-wave phonons. In forward SBS (FSBS), copropagating pump and Stokes photons are coupled by MHz-GHz frequency standing-wave phonons. Since $\omega_m << \omega_{p,S}$, we can approximate $\omega_p=\omega_S$ for the purpose of determining the optical mode shapes and wavevectors. As a result, when the pump and Stokes phonons belong to the same optical mode family (intramodal scattering), we can approximate $k_m\approx2k_p$ for BSBS and $k_m\approx 0$ for FSBS. 

Provided that all contributing effects are reversible, i.e. that each forward action has a corresponding back action, the Brillouin gain can be calculated from either the forward or back action alone \cite{wolff_stimulated_2015}. For our treatment of SBS, we use a forward action formalism closely following \cite{wiederhecker_brillouin_2019}, where we consider the scattering produced by the optical permittivity changes associated with the electro-mechanical mode (Fig.~\ref{Calculation}d). We include the well established moving boundary (MB), photoelastic (PE), and roto-optic (RO) effects in our formalism, as well as the linear electro-optic (EO) effect.

The dynamics of SBS in waveguides can be captured by the spatial power exchange between the pump and Stokes waves. We first define the optical powers $P_{p,S}$, and spatial loss rates $\alpha_{p,S}$. The steady state powers vary with distance as
\begin{align}\label{eq:G_b def}
    \partial_{z}P_p&=-G_BP_pP_S-\alpha_pP_p\\
    \partial_{z}P_S&=\pm G_BP_pP_S\mp \alpha_SP_S
\end{align}
where the $+$ $(-)$ cases correspond to forward (backward) SBS. Here we have introduced the Brillouin gain $G_B$ $(\mathrm{W}^{-1}\mathrm{m}^{-1})$, which describes the strength of the interaction. The Brillouin gain exhibits a Lorentzian dependence on the detuning between pump and Stokes frequencies $(\Delta=\omega_p-\omega_S)$ described by \cite{boyd_nonlinear_2008}
\begin{equation}
    G_B(\Delta)\propto \frac{\gamma_m^2/2}{(\gamma_m/2)^2+(\Delta-\omega_m)^2}
\end{equation}
where $\gamma_m$ is the mechanical loss rate. From here on, we assume that the interaction is phase matched and refer to the peak Brillouin gain as $G_B$. The gain is also proportional to the electromechanical quality factor $Q_m$, which is influenced by fabrication and experimental factors, so it is often more suitable to consider the normalized peak Brillouin gain $G_B/Q_m$. This gain is given by \cite{wiederhecker_brillouin_2019}
\begin{equation}\label{eq:G_b}
    G_B/Q_m=4\omega_p\frac{|\eta_{\mathrm{SBS}}|^2}{P_SP_p\mathscr{E}_m}
\end{equation}
where we introduce the linear energy density of the electromechanical wave $\mathscr{E}_m=2\omega_m^2\int\rho|\vec{u}|^2dA+2\int \vec{E}^{mech}\cdot \vec{D}^{mech} \ d\!A$. Here $\rho$ is the local density, $\vec{u}$ is the local displacement, $\vec{E}^{mech}$ and $\vec{D}^{mech}$ are the electric field and electric-displacement field associated with the electromechanical mode. The integration region $A$ is the waveguide cross section including its surroundings, and while the displacement $\vec{u}$ can be assumed to vanish in air or vacuum, $\vec{E}^{mech}$ and $\vec{D}^{mech}$ do not. The term $\eta_{\mathrm{SBS}}$ is an integral describing the overlap of the optical waves with the permittivity changes induced by the acoustic mode. Each forward action effect contributes its own permittivity change, so $\eta_{SBS}$ can be written as a sum of terms representing the moving boundary (MB), photoelastic (PE), roto-optic (RO), and electro-optic (EO) contributions:
\begin{equation}\label{eq:eta total}
    \eta_{\mathrm{SBS}}=\eta_{\mathrm{MB}}+\eta_{\mathrm{PE}}+\eta_{\mathrm{RO}}+\eta_{\mathrm{EO}} ~~.
\end{equation}

The moving boundary contribution ($\eta_{\mathrm{MB}}$) is evaluated on the boundary between materials of differing permittivity \cite{wolff_stimulated_2015}. Denoting the permittivities as $\epsilon_1$ and $\epsilon_2$, we can define their difference $\Delta\epsilon=\epsilon_1-\epsilon_2$ and a corresponding normal vector $\hat{n}$ pointing from $\epsilon_1$ to $\epsilon_2$ (Fig.~\ref{Calculation}d). The moving boundary contribution is then given by
\begin{equation}\label{eq:MB}
    \eta_{\mathrm{MB}}=\int\vec{u}^*\cdot\hat{n}\ (\Delta\epsilon\ \vec{E}^*_{p,||}\cdot\vec{E}_{S,||}-\Delta\epsilon^{-1}\ \vec{D}^*_{p,\perp}\cdot D_{S,\perp})\cdot dl
\end{equation}
where $E_\parallel$ and $D_\perp$ denote the tangential electric fields and normal electric displacement fields at the boundary, respectively, and the subscripts $p$ and $S$ denote the pump and Stokes fields. 

The photoelastic $(\eta_{\mathrm{PE}})$ and roto-optic $(\eta_\mathrm{RO})$ contributions to the gain take similar forms, since both effects relate mechanical motion to a change in the bulk permittivity as in Fig.~\ref{Calculation}d:
\begin{equation}\label{eq:bulk eta}
    \eta_{\mathrm{PE},\mathrm{RO}}=\int \vec{E}^*_p\cdot \delta \epsilon^*_{\mathrm{PE},\mathrm{RO}}(\vec{u}) \cdot \vec{E}_S \ d\!A ~~.
\end{equation}
Here, $\delta \epsilon_{\mathrm{PE},\mathrm{RO}}$ denote the change in local permittivity due the photoelastic and roto-optic effects. Specifically, the roto-optic permittivity change is given by \cite{nelson_electric_1979}
\begin{equation}\label{eq:RO}
    \delta\epsilon^{\mathrm{RO}}_{ij}=R_{ij}\epsilon_{kj}-\epsilon_{ij}R_{kj}
\end{equation}
where $R=\frac{1}{2}(\partial_j\vec{u}_i-\partial_i\vec{u}_j)$ denotes the local elastic rotation tensor and $\epsilon_{ij}=\epsilon_0n_{ij}^2$ denotes the absolute optical permittivity.

Similarly, the photoelastic tensor $p$ gives the change in inverse optical permittivity due to an applied mechanical strain $S=\frac{1}{2}(\partial_ju_i+\partial_iu_j)$ according to $\delta(\epsilon^{-1}_{\mathrm{PE}})_{ij}=p_{ijkl}S_{kl}$. From this we obtain the photoelastic permittivity change \cite{nelson_electric_1979}
\begin{equation}\label{eq:PE}
    \delta\epsilon^{\mathrm{PE}}_{ij}=-\frac{1}{\epsilon_0}\epsilon_{ik}p_{klmn}S_{mn}\epsilon_{lj} ~~.
\end{equation}
Together, \eqref{eq:MB} and \eqref{eq:bulk eta} can be used to calculate $\eta_{\mathrm{SBS}}$ in a typical treatment of SBS.

Now we must introduce a new term to account for the proposed electro-optic contribution. Conveniently, the linear electro-optic effect is defined similarly to the photoelastic effect: the linear electro-optic tensor $r$ relates an inverse permittivity change to the RF electric field $\vec{E}^{mech}$ as $\delta(\epsilon_{EO}^{-1})_{ij}=r_{ijk}\vec{E}^{mech}_{k}$. Therefore similar to the photoelastic effect, the linear electro-optic effect produces a local bulk permittivity change given by
\begin{equation}\label{eq:EO epsilon}
    \delta\epsilon^{\mathrm{EO}}_{ij}=-\frac{1}{\epsilon_0}\epsilon_{ik}r_{klm}E^{mech}_{m}\epsilon_{lj} ~~.
\end{equation}
Unlike all other contributors to SBS, this term does not exhibit an explicit dependence on displacement. Instead, it depends on the electric field associated with the electromechanical mode, which is itself determined by the full coupled equations for piezoelectricity (see Supplementary \S5.2). We can then calculate the new overlap term $\eta_{\mathrm{EO}}$ in the same manner as other bulk effects
\begin{equation}\label{eq:EO eta}
    \eta_{\mathrm{EO}}=\int \vec{E}^*_p\cdot \delta \epsilon^*_{\mathrm{EO}}(\vec{E}^{mech}) \cdot \vec{E}_S\ d\!A ~~.
\end{equation}
Equations \eqref{eq:EO epsilon} and \eqref{eq:EO eta} give us insight into the conditions for a strong electro-optic contribution. $\delta\epsilon_{EO}$ is linearly dependent on the linear electro-optic tensor $r$, so assuming the same electromechanical and optical modes, the electro-optic contribution to $G_B$ scales quadratically with $r$. Similarly, materials with large piezoelectric coefficients are desirable since $\delta\epsilon_{EO}\propto \vec{E}^{mech}$. Furthermore, the overlap integral $\eta_{EO}$ takes a similar form to $\eta_{PE}$ and $\eta_{RO}$, so like traditional optomechanical effects, the electro-optic contribution is stronger for tightly co-located optical and electromechanical modes \cite{rakich_giant_2012}. Finally, selection rules for electro-optic SBS can be formulated similarly to those for photoelastic scattering \cite{qiu_stimulated_2013, wolff_formal_2014} by replacing the strain field with $\vec{E}^{mech}$ and photoelastic tensor with $r$.

\section{Case Study: SBS in Lithium Niobate Waveguides}

Lithium Niobate (LN) is a ferroelectric material with large linear electro-optic and piezoelectric coefficients. These properties make LN important for a broad range of electro- and nonlinear optics and acoustics applications, both in bulk crystal and thin film form \cite{boes_lithium_2023}. We anticipate that these properties could also significantly influence the SBS gain through the novel EO contribution. We therefore focus our study on single-mode thin-film LN waveguides where we expect the full hybrid mode analysis to be necessary due to the strong optical and acoustic confinement.

\begin{figure}[t!]
    \begin{adjustwidth*}{0in}{0in}
    \hsize=\linewidth
    \centering
    \includegraphics[page=2,clip, trim=0.3cm 13.25cm 2cm 3cm, width=\linewidth]{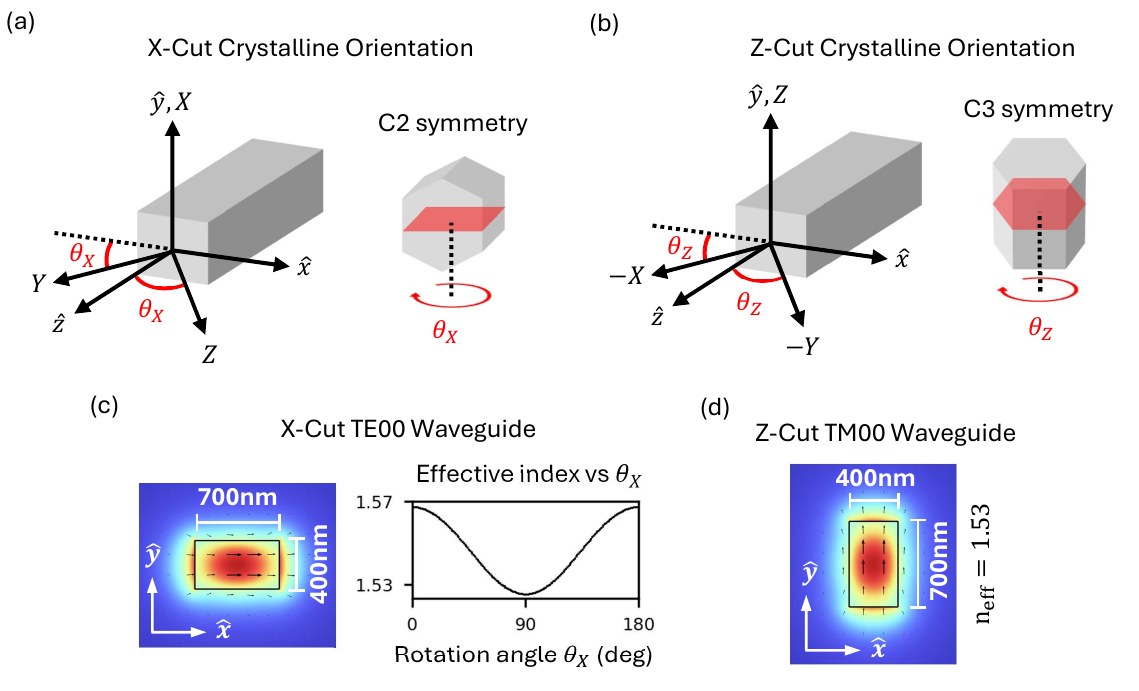}
    \caption{\textbf{LN crystalline orientations and waveguide designs. (a-b)} Crystalline axes $(X,Y,Z)$ relative to lab coordinate system $(\hat{x},\hat{y},\hat{z})$ for X and Z cut, overlaid on waveguides oriented along the $\hat{z}$ axis. X-cut LN is $C_2$ symmetric about $X$, while Z-cut is $C_3$ symmetric about $Z$, so we simulate over $\theta_X\in[0^\circ,180^\circ]$ and $\theta_Z\in[0^\circ,120^\circ]$. \textbf{(c-d)} X- and Z-cut waveguide designs. Optical intensity (color) and direction (arrows) are plotted for fundamental modes of a suspended waveguide oriented in the $\theta_{X,Z}=0^\circ$ direction at $\lambda=1550\ \mathrm{nm}$. X-cut LN is optically anisotropic, so the effective index of the X-cut waveguide varies with $\theta_X$.}
    \label{Crystalline Cuts}
    \end{adjustwidth*}
\end{figure}

Since LN is strongly anisotropic, we must guide our study by identifying which combinations of crystalline orientation and optical polarization are likely to exhibit a strong EO contribution. The strongest linear EO coefficient of LN is $r_{33}$ \cite{weis_lithium_1985}, which mixes RF and optical fields when all three waves are aligned with the crystalline Z-axis.  To take advantage of $r_{33}$, crystalline thin film LN is most commonly used in either the X-cut or Z-cut orientation. In Fig.~\ref{Crystalline Cuts}, we sketch the crystalline axes $(X,Y,Z)$ and lab axes $(\hat{x},\hat{y},\hat{z})$ for rectangular waveguides constructed with each of these cuts. As in \S 2, we orient the waveguides along the $\hat{z}$ axis. We include the rotation angles $\theta_X$ and $\theta_Z$ to describe the orientations of the two in-plane crystalline axes. X-cut LN positions the Z axis in-plane, enabling strong EO scattering between TE modes (e.g. Fig.~\ref{TE BSBS}a) near $\theta_X=90^\circ$. Z-cut LN positions the Z axis out-of-plane, enabling strong EO scattering between TM modes (e.g. Fig.~\ref{TM BSBS}a). These cuts also exhibit different symmetries. Since X-cut LN is C2 symmetric about the $X$ axis (Fig.~\ref{Crystalline Cuts}a), we only need to consider rotation angles $\theta_X\in[0^\circ,180^\circ]$ to capture the full range of possible behavior. Z-cut LN is C3 symmetric about $Z$ (Fig.~\ref{Crystalline Cuts}b), so we only consider $\theta_Z\in[0^\circ,120^\circ]$.

With these arrangements, we can compute the SBS gain in LN waveguides using our updated theoretical formulation above and explore the nature of the electro-optic contribution. We perform finite element simulations for the optical and electromechanical modes using COMSOL Multiphysics. To understand how each scattering effect contributes to SBS, we calculate the normalized Brillouin gain (Eqn. \ref{eq:G_b}) assuming that only each individual effect contributes. For example, we calculate $G_{B}^{\mathrm{EO}}/Q_m$ by replacing eq.~\ref{eq:eta total} with $\eta_{\mathrm{SBS}}=\eta_{\mathrm{EO}}$. We also calculate a gain including all of the acousto-optic scattering effects but not the EO effect, i.e. $\eta_{\mathrm{SBS}}=\eta_{\mathrm{PE}}+\eta_{\mathrm{MB}}+\eta_{\mathrm{RO}}$. Finally, we calculate $G_{B}/Q_m$ including all effects according to the updated formalism in eq.~\ref{eq:eta total}. We perform all optical simulations at 1500 nm optical wavelength, and include details on the material coefficients in the supplement.

\subsection{Backward SBS in Suspended X-cut LN}

\begin{figure}[htbp]
    \begin{adjustwidth*}{-0.5in}{-0.5in}
    \hsize=\linewidth
    \centering
    \includegraphics[page=3,clip, trim=0.2cm 8.5cm 0.5cm 3cm, width=\linewidth]{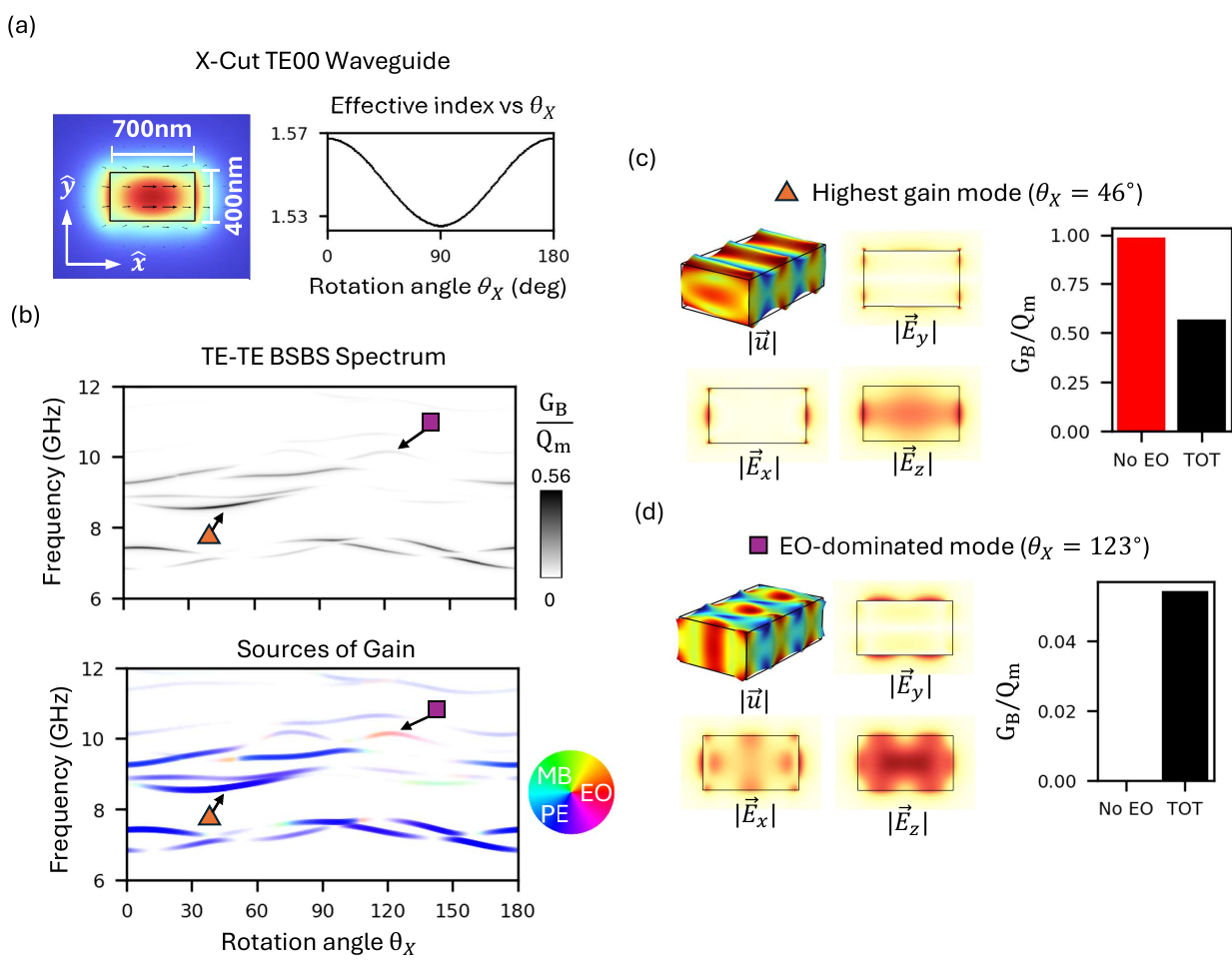}
    \caption{\textbf{Suspended X-cut waveguide, TE-TE BSBS (a)} Suspended waveguide dimensions, TE00 mode profile ($\theta_X=0$), and effective index vs $\theta_X$. \textbf{(b)} Backward Brillouin spectrum assuming $Q_m=300$ and gain source visualization. The spectrum is primarily dominated by the photoelastic (PE) contribution, though some weakly Brillouin-active modes are dominated by the electro-optic (EO) and moving-boundary (MB) contributions. We identify the mode with the strongest Brillouin gain (triangle symbol) and a mode where the EO effect plays a strong role (square symbol). \textbf{(c)} Displacement and electric fields associated with the strongest gain mode, alongside visualizations of the mechanical-only and total gains. The EO contribution reduces the gain by $\sim 42\%$. \textbf{(d)} Same as \textbf{(c)}, but for the EO-dominated mode. The EO contribution accounts for effectively all of the SBS gain, although the total gain is small.}
    \label{TE BSBS}
    \end{adjustwidth*}
\end{figure}

We first consider suspended waveguides, which are able to host guided acoustic modes satisfying the phase-matching conditions for both BSBS and FSBS \cite{rakich_giant_2012}. We set up a rectangular X-cut waveguide that hosts a strongly confined TE00 mode (Fig.~\ref{TE BSBS}a). We next simulate the electromechanical modes this waveguide supports at each angle $\theta_X\in[0^\circ,180^\circ]$, narrowing our search to only modes that satisfy the phase matching condition for BSBS ($k_m=2k_p$). Finally, we calculate the normalized Brillouin gain for each of the simulated electromechanical modes at each angle. In Fig.~\ref{TE BSBS}b, we plot the normalized BSBS spectrum as a function of $\theta_X$. Here and for all other Brillouin spectra we plot, we arbitrarily assume a quality factor of 300 for all mechanical modes. Since we consider the normalized Brillouin gain, this choice only effects the Brillouin linewidths and is made for legibility. The spectrum is confirmed to be highly angle dependent, with a peak gain of $G_B/Q_m=0.56\mathrm{W}^{-1}\ \mathrm{m}^{-1}$.

We also develop a scatterplot visualization of how the different scattering effects combined with their back-action counterparts contribute to the gain (e.g. Fig.~\ref{TE BSBS}b, bottom). We assign each simulated mode-angle combination an RGB color vector $(G_B^{EO},G_B^{MB},G_B^{PE})$ mapping the EO-, MB-, and PE-only Brillouin gains to red, green, and blue, respectively. We then normalize the vector such that the largest individual-source gain is set to the maximal saturation value. Here and for the rest of our analysis, we do not explicitly plot the roto-optic-only gain since it has a negligible impact on total gain, though we include it in the total gain calculation. We also set the opacity of each point to be proportional to the total Brillouin gain, normalized such that the mode with the strongest gain is fully opaque. By this approach, in Fig.~\ref{TE BSBS}b, we produce the gain source visualization for the BSBS spectrum. We can readily identify most of the modes in the spectrum as PE-dominated (blue regions), although some weaker gain modes do appear as EO-dominated (faint red regions) and MB-dominated (faint green regions).

To investigate the effect of the EO contribution, we identify the mode with the strongest total gain (Fig.~\ref{TE BSBS}b, triangle symbol), and also a strongly EO-dominated mode (Fig.~\ref{TE BSBS}c, square symbol). In Fig.~\ref{TE BSBS}c, we plot the normalized Brillouin gain associated with the highest gain mode both ignoring and including the EO contribution, and find that the EO contribution in fact reduces the Brillouin gain by $\sim42\%$ through destructive interference. We also plot the mechanical displacement and electric fields associated with the highest gain mode for visualization purposes. Interestingly, the $E$ field associated with this electromechanical mode is almost entirely $\hat{z}$ polarized inside the waveguide, while the optical $E$ field (Fig.~\ref{TE BSBS}a) is mostly $\hat{x}$ polarized. Since $r_{33}$ mixes waves of the same polarization, this suggests that a far greater EO contribution might be attainable through a phase-matched electro-mechanical mode with a strong $E_{\hat{x}}$ field.

We plot the same information for the strongly EO dominated mode (Fig.~\ref{TE BSBS}d), and find that while the mode exhibits a modest total gain, it is effectively Brillouin-inactive without the EO contribution. We refer to this surprising case as pure stimulated EO scattering (SES), since although a mechanical deformation is present, the SBS effect is entirely mediated by interactions between RF and optical frequency electric fields. Again, this mode exhibits a mostly $\hat{z}$ polarized $E$ field within the waveguide, suggesting that better overlap of the RF and optical $E$ fields could dramatically enhance the EO contribution.

\begin{figure}[htbp]
    \begin{adjustwidth*}{-0.5in}{-0.5in}
    \hsize=\linewidth
    \centering
    \includegraphics[page=4,clip, trim=0.1cm 3cm 0cm 3cm, width=\linewidth]{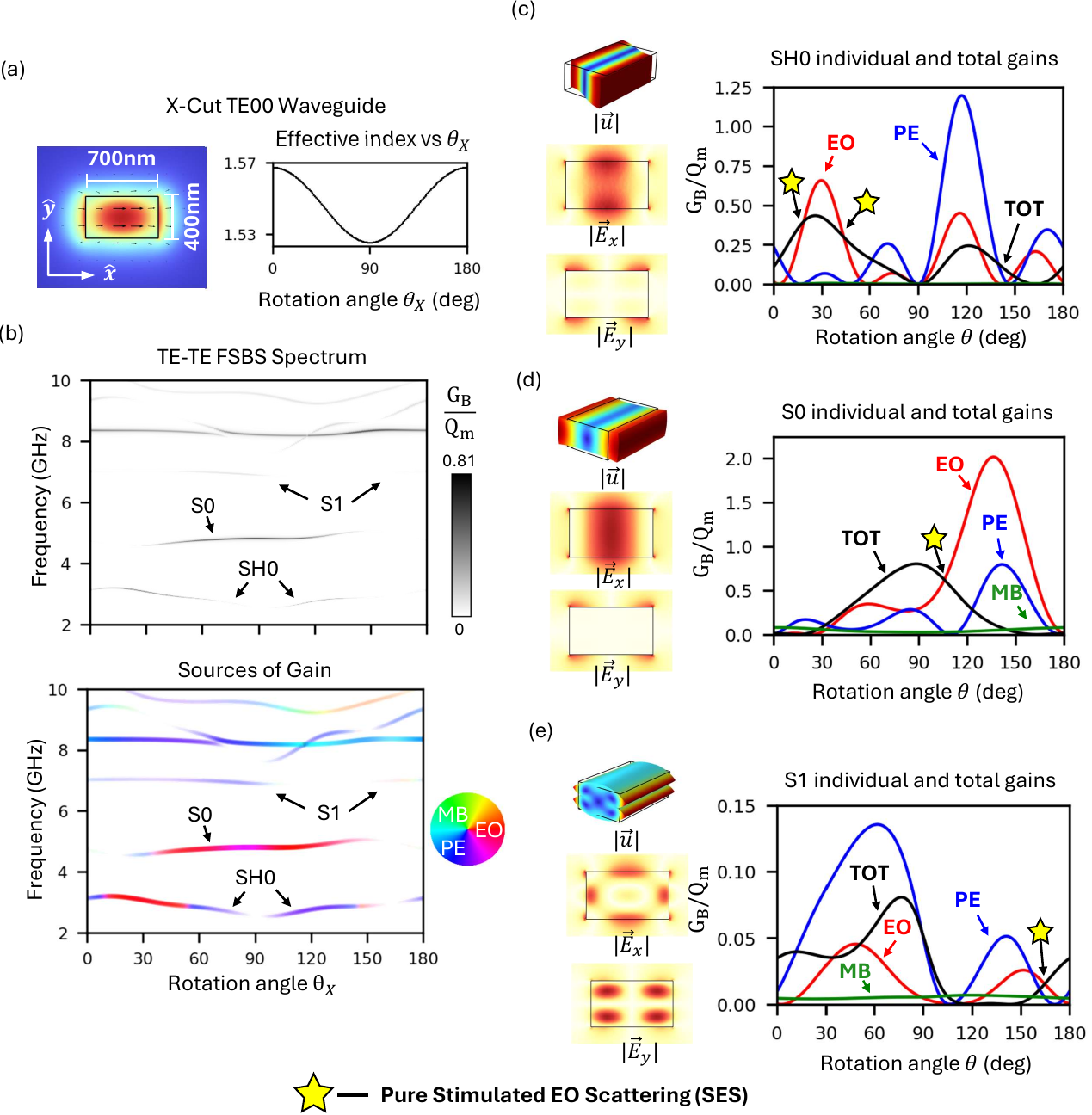}
    \caption{\textbf{Suspended X-cut waveguide, TE-TE FSBS (a)} Suspended waveguide dimensions, TE00 mode profile ($\theta_X=0$), and effective index vs $\theta_X$. \textbf{(b)} Forward Brillouin spectrum assuming $Q_m=300$ and gain source visualization. The spectrum shows a mix of EO, MB, and PE contributions, with several high gain modes dominated by the EO contribution. We identify the first three FSBS active modes as the fundamental width-shear mode (SH0), fundamental symmetric or ``breathing'' mode (S0), and first order symmetric mode (S1). \textbf{(c-e)} Mechanical and electrical modes shapes for the first three modes, alongside graphs of the normalized individual-source and total Brillouin gains for each mode vs $\theta_X$. Each mode features at least one incidence of pure stimulated EO scattering, indicated by star symbols.}
    \label{TE FSBS}
    \end{adjustwidth*}
\end{figure}

\subsection{Forward SBS in Suspended X-cut LN}

We next consider the forward scattering case in Fig.~\ref{TE FSBS}. We again plot the Brillouin spectrum assuming $Q_m=300$ and the gain source visualization (Fig.~\ref{TE FSBS}b), and this time we find several EO-dominated modes with large gains. This is strikingly different from the BSBS case (Fig.~\ref{TE BSBS}), in which the EO contribution only dominated the gain for very weakly Brillouin active modes. The Brillouin active modes are also more spectrally separated and resolvable than the BSBS case, which allows us to track the evolution of the first three FSBS active modes across the full angle range (Fig.~\ref{TE FSBS} c-e). We also plot the displacement and $E$ fields associated with the first three electromechanical modes at $\theta_X=0$ (Fig.~\ref{TE FSBS} c-e), which helps us identify them as the fundamental width-shear mode (SH0, Fig.~\ref{TE FSBS}c), fundamental transverse symmetric or ``breathing" mode (S0, Fig.~\ref{TE FSBS}d), and the first order transverse symmetric mode (S1, Fig.~\ref{TE FSBS}e). In contrast with the BSBS modes we considered in Fig.~\ref{TE BSBS}, the electromechanical $E$ fields here feature a strong $E_{\hat{x}}$ component inside the waveguide. This explains the larger EO contributions in FSBS, since the optical and electromechanical $E$ fields are well aligned to interact through $r_{33}$ in LN. While transverse symmetric modes have been extensively utilized for on-chip FSBS \cite{rakich_giant_2012, shin_tailorable_2013, qiu_stimulated_2013, kittlaus_large_2016, van_laer_interaction_2015, zhou_electrically_2024, wang_demonstration_2021, shin_control_2015, lei_anti-resonant_2024}, the width-shear mode is longitudinal and would not usually be considered for FSBS. In LN, however, the $p_{14}$ photoelastic tensor component is nonzero \cite{weis_lithium_1985}, enabling TE-TE photoelastic scattering through longitudinal shear motion, and TE-TE electro-optic scattering is enabled by the strong $E_{\hat{x}}$ field.

To better understand how the gain sources interact, in Fig.~\ref{TE FSBS}c-e we ignore mode frequency and plot the individual-source and total gains for the SH0, S0, and S1 modes. For some angles, we find that the total gain is smaller (larger) than the EO- or PE-only gain, since the effects act destructively (constructively). Once again we find that pure SES occurs at some angles for all of these modes. A convenient shorthand to identify SES is to identify points where the EO-only gain is equal to the total Brillouin gain while all other individual-source gains are smaller. In other words, pure SES appears when all of the mechanical scattering effects collectively cancel out or equal zero ($\eta_{MB}+\eta_{PE}+\eta_{RO}=0$), while EO scattering still occurs. In contrast with the BSBS case of pure SES (Fig.~\ref{TE BSBS}d), several of these SES modes have large Brillouin gains relative to the rest of the spectrum, with the strongest pure SES mode exhibiting $G_B/Q_m=0.6\ \mathrm{W}^{-1}\mathrm{m}^{-1}$. We therefore predict that large SBS gains may be achievable through exclusively electro-optic interactions.

\subsection{Backward and Forward SBS in Suspended Z-cut LN}

\begin{figure}[b!]
    \begin{adjustwidth*}{-0.5in}{-0.5in}
    \hsize=\linewidth
    \centering
    \includegraphics[page=5,clip, trim=0.1cm 11cm 0cm 3cm, width=\linewidth]{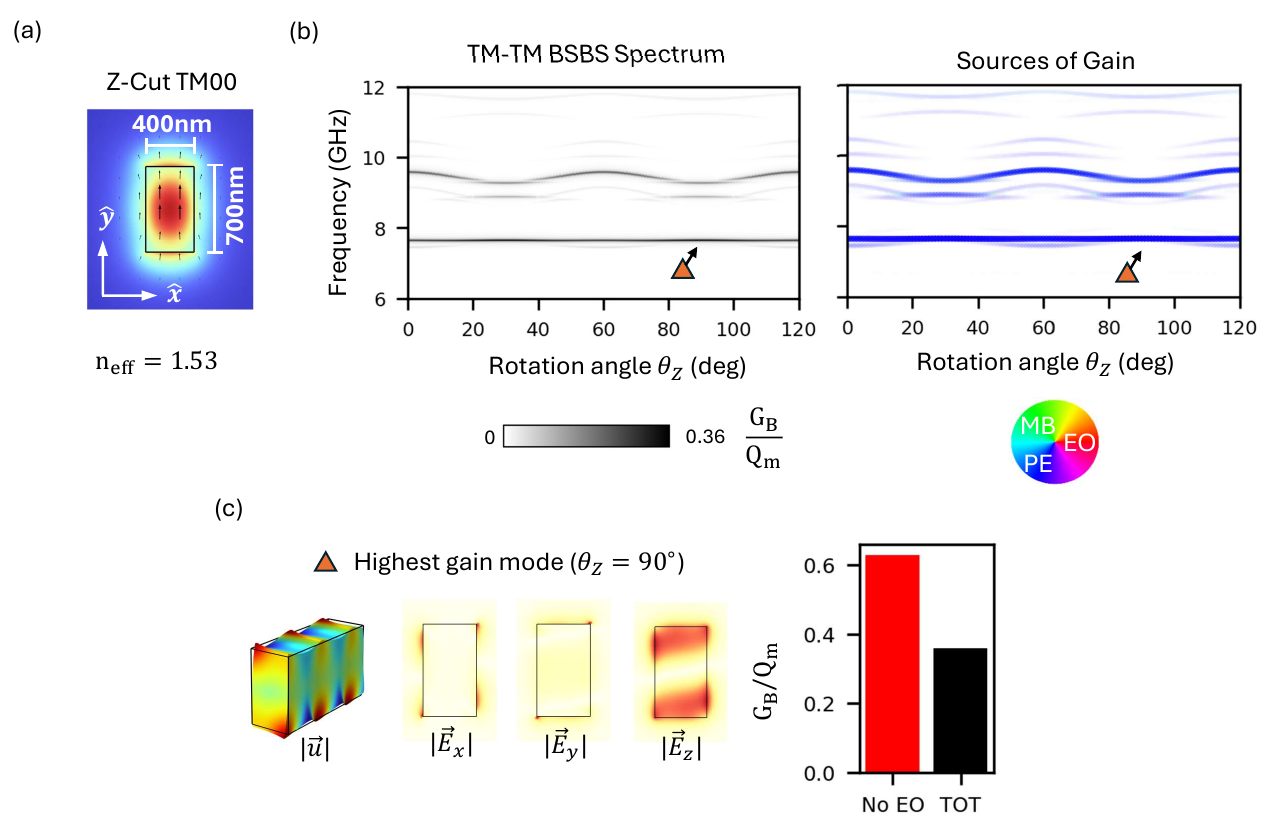}
    \caption{\textbf{Suspended Z-cut waveguide, TM-TM BSBS (a)} Suspended waveguide dimensions, TM00 mode profile ($\theta_Z=0^\circ$), and effective index. \textbf{(b)} Backward Brillouin spectrum assuming $Q_m=300$ and gain source visualization. The spectrum appears entirely dominated by the photoelastic (PE) contribution. We identify the mode with the strongest Brillouin gain (triangle symbol). \textbf{(c)} Displacement and electric fields associated with the strongest gain mode, alongside visualizations of the mechanical-only and total gains. The EO contribution reduces the gain by $\sim 42\%$.}
    \label{TM BSBS}
    \end{adjustwidth*}
\end{figure}

\begin{figure}[htbp]
    \begin{adjustwidth*}{-0.5in}{-0.5in}
    \hsize=\linewidth
    \centering
    \includegraphics[page=6,clip, trim=0.1cm 6.5cm 0cm 3cm, width=\linewidth]{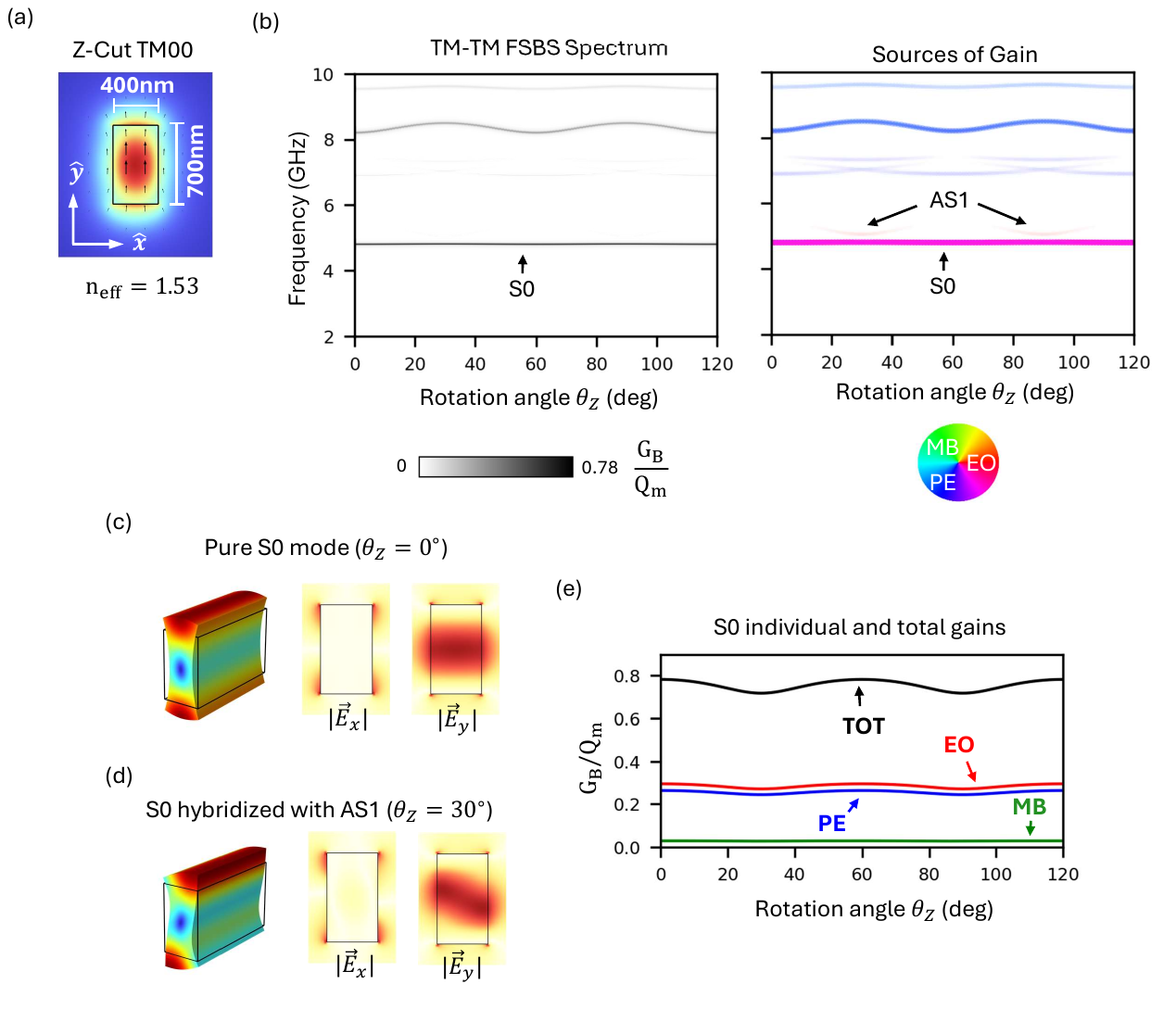}
    \caption{\textbf{Suspended Z-cut waveguide, TM-TM FSBS (a)} Suspended waveguide dimensions and TM00 mode profile ($\theta_Z=0^\circ$). \textbf{(b)} Forward Brillouin spectrum assuming $Q_m=300$, alongside a visualization of the dominant gain source for each simulated mode. The spectrum is primarily PE-dominated, but one mode with strong PE and EO contributions emerges. We identify this mode as S0, although it exhibits some hybridization with the first-order antisymmetric mode AS1 near $\theta_Z=30^\circ$ and $90^\circ$, as evidenced by an avoided crossing where AS1 becomes slightly Brillouin active due to the hybridization. \textbf{(c)} Mechanical and electric mode shapes for S0 at $\theta_Z=0^\circ$, where hybridization with AS1 is minimal. \textbf{(d)} Same as \textbf{(c)}, but at $\theta_Z=30^\circ$, where S0 strongly hybridizes with S1. \textbf{(e)} Plot of normalized individual-source and total Brillouin gains for the S0 mode vs $\theta_X$. No instances of SES appear, but constructive interference between the EO and PE effects result in a significantly enhanced gain versus if the EO contribution were not considered.}
    \label{TM FSBS}
    \end{adjustwidth*}
\end{figure}

We next explore TM-TM FSBS and BSBS in suspended Z-cut waveguides. For the sake of comparison, we use the same waveguide dimensions as the X-cut case (Fig.~\ref{TE BSBS}a), but align the longer side with the $\hat{y}$ axis so that TM modes experience a similar level of confinement (Fig.~\ref{TM BSBS}). Unlike the X-cut case, the effective index does not vary with angle since the permittivity tensor of LN is unaffected by rotations about the crystalline Z axis \cite{weis_lithium_1985}. We again begin with BSBS, and plot the Brillouin spectrum and gain source visualization in Fig.~\ref{TM BSBS}b. This time we find that the spectrum appears entirely PE dominated (Fig.~\ref{TM BSBS}b). We identify the mode with the largest gain (Fig.~\ref{TM BSBS}b, triangle symbol), and plot its associated displacement and $E$ fields (Fig.~\ref{TM BSBS}c). We also plot the normalized Brillouin gain both ignoring and including the EO contribution (Fig.~\ref{TM BSBS}c), and find that the EO contribution again reduces the gain by $\sim 42\%$. Similar to the TE-TE BSBS mode in Fig.~\ref{TE BSBS}c, the mostly $\hat{y}$-polarized optical and mostly $\hat{z}$-polarized electro-mechanical $E$ fields are not well aligned for a strong EO contribution through $r_{33}$.

In the FSBS case, we find that the EO contribution again plays a much larger role. Fig.~\ref{TM FSBS} presents we plot the FSBS spectrum and gain source visualization and we find that the most strongly Brillouin active mode features strong EO and PE contributions. We identify this mode as S0, although near $\theta_Z=30^\circ$ and $90^\circ$, there is an avoided crossing with the first order antisymmetric transverse mode (AS1). The AS1 mode is only faintly visible in the spectrum near the avoided crossings, indicating hybridization with S0. In Fig.~\ref{TM FSBS} c-d, we plot the mode shapes of S0 both at $\theta_Z=0^\circ$ (c) and $\theta_Z=30^\circ$ (d), confirming that S0 and AS1 mix near the crossings. Similar to the TE FSBS case, we find that that S0 features a strong $E_{\hat{y}}$ field inside the waveguide, which is necessary for strong EO interactions through $r_{33}$ in Z-cut LN. Finally, we plot the individual-source and total gains for the S0 mode (Fig.~\ref{TM FSBS}e). Although SES does not occur in this specific example, we again confirm the importance of including the EO contribution in Brillouin gain calculations, since the gain would otherwise be significantly under-predicted at all angles.

\subsection{Backward SBS in LN Strip Waveguides}

\begin{figure}[htbp]
    \begin{adjustwidth*}{-0.5in}{-0.5in}
    \hsize=\linewidth
    \centering
    \includegraphics[page=7,clip, trim=0.1cm 5cm 0cm 3cm, width=\linewidth]{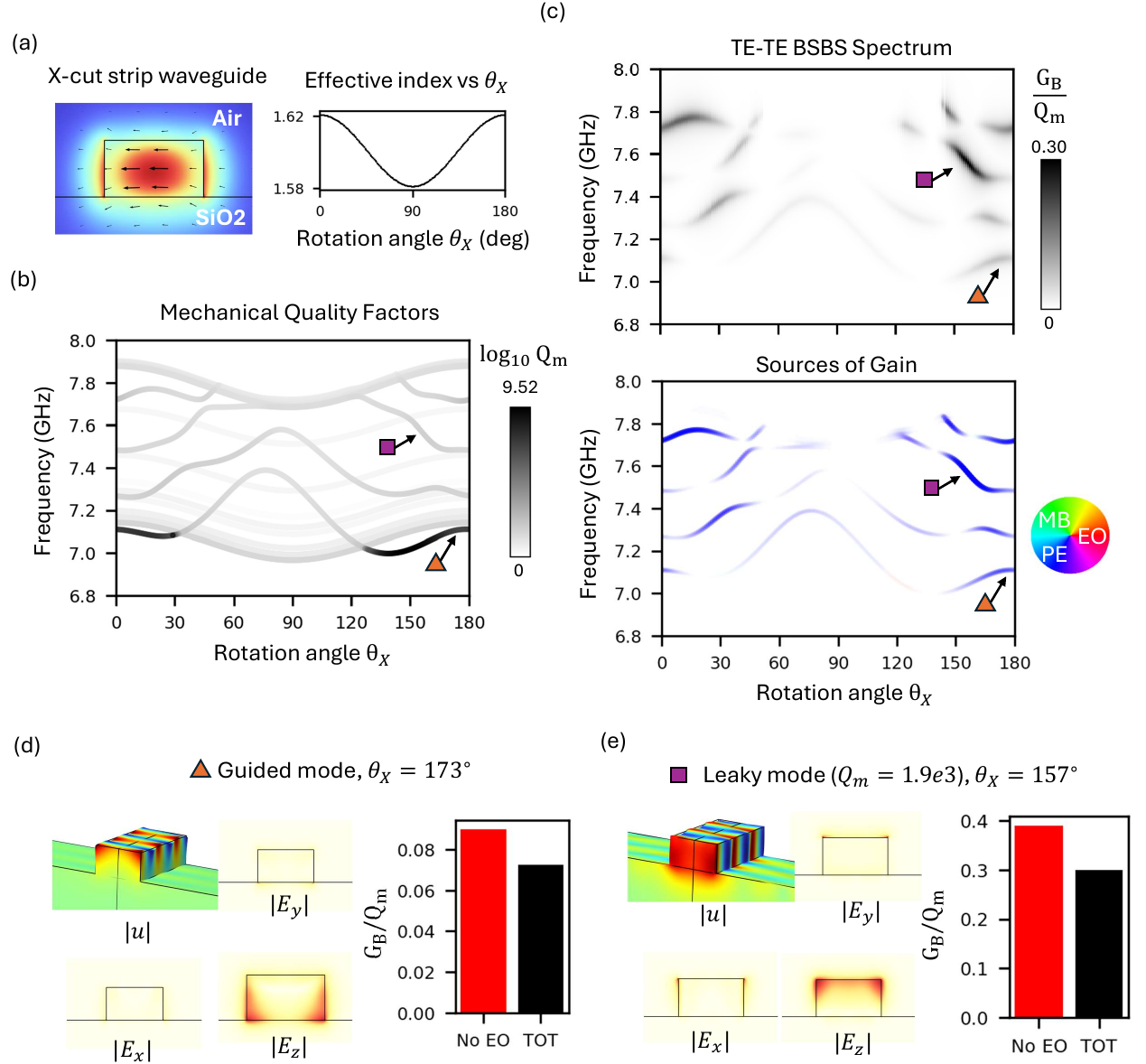}
    \caption{\textbf{X-cut strip waveguide, TE-TE BSBS (a)} TE00 mode profile ($\theta_X=0^\circ$) and effective index vs $\theta_X$ for an identical LN waveguide to those in Figs \ref{TE BSBS} and \ref{TE FSBS}, but with a SiO2 substrate. \textbf{(b)} Quality factors $Q_m$ of the electro-mechanical eigenmodes versus $\theta_X$. Full acoustic guidance is only possible at some angles, though leaky modes at least partially located in the LN are also visible. We can identify these modes by their lower symmetry relative to the SiO2-confined modes, which exhibit the same reflection symmetry about $\theta_X=90^\circ$ as the optical effective index. \textbf{(c)} Backward Brillouin spectrum assuming $Q_m=300$ and gain source visualization. We identify the most strongly Brillouin-active guided mode (triangle symbol), as well as the most strongly Brillouin active mode overall (square symbol), which is leaky. The spectrum appears entirely dominated by the PE contribution. \textbf{(d)} Displacement and $E$ fields associated with the most strongly Brillouin active guided mode, alongside visualizations of the mechanical-only and total gains. The EO contribution reduces the gain by $\sim 20\%$. \textbf{(e)} same as \textbf{(d)}, but for the most strongly Brillouin active leaky mode, the EO contribution reduces the gain by $\sim 23\%$.}
    \label{TE Strip}
    \end{adjustwidth*}
\end{figure}

While suspended waveguides offer advantages such as enhanced mode confinement and FSBS activity, they are also difficult to fabricate and exhibit poor power handling. By adding a substrate beneath our waveguide designs, we can ameliorate these issues at the cost of decreased mode confinement and effectively eliminating the FSBS gain. For these reasons, surface-acoustic-wave (SAW) based SBS \cite{neijts_-chip_2024} has attracted significant recent attention and accounts for many of the theoretical treatments \cite{rodrigues_stimulated_2023} and experimental realizations of SBS on TFLN \cite{yang_stimulated_2023, ye_integrated_2025, rodrigues_cross-polarized_2025}. For the sake of direct comparison, we simulate the same X and Z cut waveguide designs as as in previous sections with the addition of an $\mathrm{SiO_2}$ substrate, which is referred to as a strip waveguide design.

\begin{figure}[htbp]   
    \begin{adjustwidth*}{-0.5in}{-0.5in}
    \hsize=\linewidth
    \centering
    \includegraphics[page=8,clip, trim=0.1cm 4cm 0.5cm 3cm, width=\linewidth]{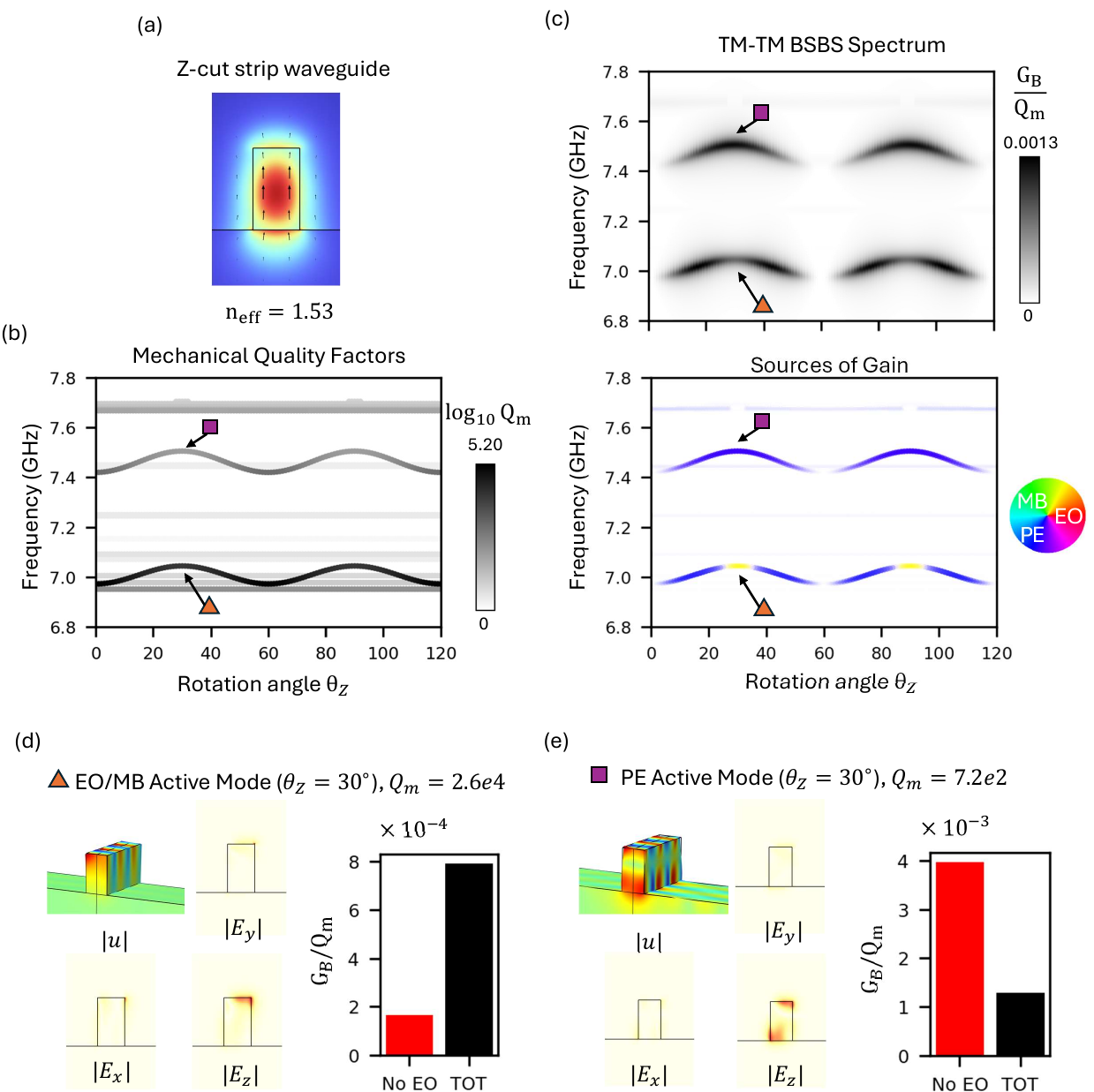}
    \caption{\textbf{Z-cut strip waveguide, TM-TM BSBS (a)} TM00 mode profile ($\theta_Z=0^\circ$) and effective index for an identical LN waveguide to those in Figs \ref{TM BSBS} and \ref{TM FSBS}, but with a SiO2 substrate. \textbf{(b)} Quality factors $Q_m$ of the electro-mechanical eigenmodes versus $\theta_Z$. All modes are at least slightly leaky, with modes partially located in the LN identifiable by their non-flat angular dispersion. \textbf{(c)} Backwards Brillouin spectrum assuming $Q_m=300$ and gain source visualization. The entire spectrum features weak gains compared to the other spectra we have considered. We identify the most strongly Brillouin active mode (square symbol) as well as a mode dominated by a combination of EO and MB contributions (triangle symbol). \textbf{(d)} Displacement and $E$ fields associated with the EO and MB active mode, alongside visualizations of the mechanical-only and total gains. \textbf{(e)} Same as \textbf{(d)}, for the other most strongly Brillouin active mode we identified. In both modes, the inclusion of the EO contribution causes a large relative change in the Brillouin gain, but the absolute change is very small.}
    \label{TM Strip}
    \end{adjustwidth*}
\end{figure}

In Fig.~\ref{TE Strip}, we consider TE-TE SBS in the strip waveguide counterpart to our X-cut waveguide. We again simulate the TE00 mode over $\theta_X\in[0^\circ,180^\circ]$, and find that the effective index is increased relative to the suspended case (Fig.~\ref{TE Strip}a) due to the the introduction of a substrate. We plot the TE00 mode at $\theta_X=0^\circ$ (Fig.~\ref{TE Strip}a) and confirm that some of the optical energy leaks into the oxide, indicating a lower degree of optical confinement in the LN. The introduction of a substrate also provides a route for acoustic leakage, which leads to non-negligible electromechanical loss rates in our simulations. In Fig.~\ref{TE Strip}b, we plot the simulated quality factors $Q_m$ of the electromechanical modes satisfying the phase matching condition for BSBS. The waveguide only supports fully guided BSBS modes at some angles, which we can identify both by their large quality factor and because they occur at a lower frequency than the lowest modes located in the oxide. We can also identify leaky modes partly located in the LN based on the symmetry of their angular dispersion. While primarily $\mathrm{SiO_2}$-guided modes posses roughly the same reflection symmetry about $\theta_X=90^\circ$ as the optical effective index, modes located in the LN break this symmetry due to the mechanical anisotropy of LN. The dense cluster of oxide modes near the top of the plot sets a ceiling on the frequency range we can simulate, since the computational cost of tracking LN-confined modes among dense spurious modes becomes high.

In Fig.~\ref{TE Strip}c, we plot the BSBS spectrum and gain-source visualization for the X-cut strip waveguide. We continue to assume $Q_m=300$ for the strip waveguide Brillouin spectra, since the Q factors that were simulated in Fig.~\ref{TE Strip}b account only for acoustic leakage and not for other sources of loss. While the peak normalized Brillouin gain is nearly as large as the suspended BSBS case (Fig.~\ref{TE BSBS}b), this gain is achieved by a leaky mode (Fig.~\ref{TE Strip}c, triangle symbol) while the most strongly Brillouin active guided mode (Fig.~\ref{TE Strip}c, square symbol) has a lower gain. Unlike the suspended case, the entire Brillouin spectrum is dominated by the PE contribution. We plot the displacement and $E$ fields for the strongest guided (Fig.~\ref{TE Strip}d) and leaky (Fig.~\ref{TE Strip}e) modes, alongside visualizations of gain both including and ignoring the EO contribution. The EO contribution acts destructively in both cases, decreasing the gain by $\sim 20\%$ for the guided mode and $\sim 23\%$ for the leaky mode. Similar to the suspended BSBS cases we examined in Fig.~\ref{TE BSBS} and Fig.~\ref{TM BSBS}, both of these modes feature mostly $\hat{z}$ polarized $E$ fields within the waveguide. As a result, these modes engage the $r_{33}$ coefficient of LN less strongly than the strongly EO-dominated FSBS modes we considered in Fig.~\ref{TE FSBS} and Fig.~\ref{TM FSBS}, which explains the relatively weaker EO contribution.

In Fig.~\ref{TM Strip}, we consider the strip counterpart to our Z-cut waveguide for TM-TM scattering. Similar to the X-cut case, the effective index is increased compared to the suspended case as the mode leaks into the oxide (Fig.~\ref{TM Strip}a). We again plot the Q factors of the electro-mechanical modes (Fig.~\ref{TM Strip}b), and find that although no modes are completely guided, some do feature very low loss ($Q_m>10^5$). Similar to the X-cut case, we can identify modes located in the LN by their broken symmetry. Since the optical effective index does not vary with $\theta_Z$, modes located in the $\mathrm{SiO_2}$ have a flat angular dispersion, while the frequencies of LN-localized modes vary with angle.

We again plot the Brillouin spectrum and gain source visualization (Fig.~\ref{TM Strip}c). Two modes exhibit the strongest Brillouin activity, although the entire Brillouin spectrum is extremely weak compared to the other waveguides we examine. We identify the strongest gain mode, which is PE-dominated (square symbol) as well as an EO- and MB-dominated mode (triangle symbol). Both of these modes exhibit a very large proportional change in gain depending whether the EO contribution is included or excluded (Fig.~\ref{TM Strip} d,e).

\subsection{Pure SES in a suspended ridge waveguide}

\begin{figure}[htbp]   
    \begin{adjustwidth*}{-0.5in}{-0.5in}
    \hsize=\linewidth
    \centering
    \includegraphics[page=9,clip, trim=0.1cm 4cm 0cm 3cm, width=\linewidth]{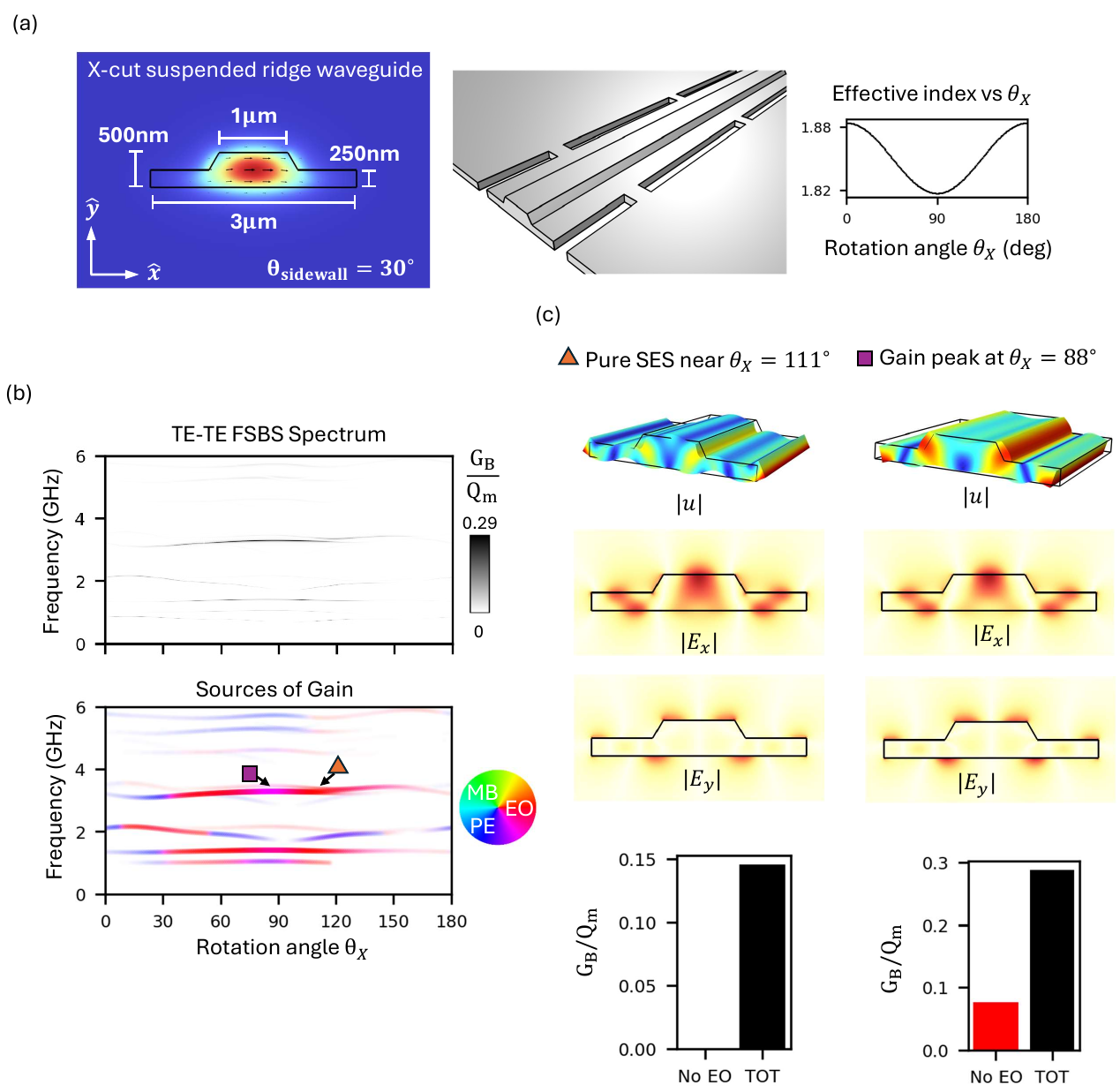}
    \caption{\textbf{X-cut suspended ridge waveguide, TE-TE FSBS (a)} Suspended waveguide dimensions, TE00 mode profile ($\theta_X=0$), and effective index vs $\theta_X$. \textbf{(b)} Forward Brillouin spectrum assuming $Q_m=300$ and gain source visualization. The spectrum shows a mix of EO, MB, and PE contributions, with several high gain modes dominated by the EO contribution. We identify a case of pure SES (triangle symbol) and the mode with the strongest gain (square symbol). \textbf{(c)} Displacement and $E$ fields associated with the pure SES mode and strongest gain mode, alongside visualizations of the mechanical-only and total gains.}
    \label{Suspended Ridge}
    \end{adjustwidth*}
\end{figure}

In our exploration of SBS in TFLN, we have found that suspended waveguides host FSBS modes exhibiting large EO-dominated Brillouin gains. In particular, we have found several cases of high gain forward SES in suspended X-cut waveguides. Unfortunately, fabrication of spatially-invariant rectangular suspended waveguides is not possible, since the waveguide must somehow be supported. One solution is to fabricate a suspended ridge waveguide supported by tethers (Fig.~\ref{Suspended Ridge}a) \cite{qiu_stimulated_2013, kittlaus_large_2016, kittlaus_-chip_2017, otterstrom_silicon_2018}, which has been implemented in TFLN \cite{yu_-chip_2025} though FSBS has not been reported. In this section, we modify our idealized X-cut suspended waveguide to a more practical suspended ridge design and propose an experimentally feasible realization of high-gain stimulated electro-optic scattering.

We identify the dimensions of our ridge waveguide design in Fig.~\ref{Suspended Ridge}a. This structure can be realized by a two etch process in which the waveguide is defined by a partial etch, and then long slots are etched through the full thickness of the film so that the underlying substrate can be removed. We expect many factors to contribute non-trivially to the total gain of a suspended LN ridge waveguide. For example, although decreasing the waveguide top width increases optical confinement, it may also reduce the proportion of the acoustic energy located in the waveguide and require a larger total structure width to prevent the optical field from interacting with the tethers. We leave structural optimizations to future studies.

In Fig.~\ref{Suspended Ridge}b, we plot the FSBS spectrum and gain visualization for the waveguide. Similar to in the idealized rectangular waveguide, several EO-dominated modes with large gains appear. We identify the strongest gain mode (square symbol), and yet another incidence high-gain pure SES (triangle symbol), which appear on the same angular dispersion branch at different angles. Unlike the idealized rectangular waveguide (Fig.~\ref{TE FSBS}), we do not track the evolution of modes with varying angles, since the spectrum is crowded and features numerous avoided crossings. In Fig.~\ref{Suspended Ridge}c, we plot the displacement and $E$ fields associated with the high gain and pure SES electromechanical modes, and find that they are primarily characterized by symmetric transverse motion. Additionally, both feature strong $E_x$ fields inside the waveguide, enabling EO scattering through $r_{33}$. Both electromechanical modes also feature strong SBS gains when the EO contribution is included and large deviations from the mechanical-only theory, suggesting that such a waveguide design could be used to experimentally verify the EO contribution to SBS.

\section{Discussion}

In this work we have developed a full modal analysis of Brillouin in piezoelectric materials bringing the electro-optic and $\chi^{(2)}$ effects on the same footing as well-known acousto-optic couplings. We confirm that these effects can substantially modify the gain in nanophotonic waveguides through interference with other forward and back action pairs. Importantly, our results show that a proper inclusion of the EO contribution is critical for developing SBS-active integrated photonics on thin film piezoelectrics like LN \cite{rodrigues_stimulated_2023, luo_visible_2025, haerteis_suspended_2025, rodrigues_cross-polarized_2025, yang_stimulated_2023, li_-chip_2026, yu_-chip_2025, ye_integrated_2025}. Far larger EO-enhanced gains might also be achievable in next-generation electro-optic materials such as Barium Titanate (max clamped Pockels coefficient $r_{42}=720$ pm/V \cite{zgonik_dielectric_1994}) and cryogenic Strontium Titanate (max unclamped Pockels coefficient $r_{33}=1060$ pm/V \cite{anderson_quantum_2025}, clamped values not reported). Similarly, heterostructures combining the individual advantages of strong piezoelectric and electro-optic effects in different materials could be developed.

We find that we must also pay special attention to the emergence of pure stimulated electro-optic scattering (SES), a novel nonlinear optical phenomenon in which the Brillouin gain originates entirely from couplings between optical and RF electric fields. Generally, electro-optic equivalents to SBS are considered forbidden due to the difficulty with phase matching between these modes. However, since the hybrid electro-mechanical modes involved in SES propagate at acoustic velocities, the all-electric interaction to contribution is phase-matched and can contribute to Brillouin gain. This marks a significant departure from state-of-the-art theory.

Just over a decade ago, the introduction of radiation pressure and moving-boundary effects \cite{rakich_giant_2012} to SBS theory significantly expanded the design space for enhancing and controlling the SBS gain \cite{rakich_giant_2012,dostart_giant_2015}. The novel EO contribution we have identified in this work has the potential to further expand the ability to control and harness SBS independent of index contrast and the photoelastic tensor, opening a new pathway for improving Brillouin-based devices in microwave and quantum photonics.

{\footnotesize \putbib[references]}

\section*{Acknowledgments and Funding}

We thank Oğulcan Örsel for helpful comments and discussions. This work was sponsored by the Defense Advanced Research Projects Agency (DARPA) under Cooperative Agreement D24AC00003, the US Office of Naval Research (ONR) Multi-University Research Initiative grant N00014-20-1-2325, and the Army Research Office (ARO) grant W911NF-23-1-0219. The views and conclusions contained herein are those of the authors and should not be interpreted as necessarily representing the official policies or endorsements, either expressed or implied, of DARPA, ONR, ARO, or the US Government.

\title{Supplementary Information:\\ Stimulated Electro-optic Scattering}

\author{}

\date{}

\maketitle{}

\section{Lithium niobate material properties}

To simulate the Brillouin gain in lithium niobate (LN), we require the tensors for its optical, mechanical, and microwave/RF properties, as well as for the various effects which couple these regimes.

\subsection{Optical properties}

The relative optical permittivity of LN is directly tied to the refractive index, and at 1550nm is given by
\begin{equation}
    \epsilon=\begin{pmatrix}
        \epsilon_{11} & 0 & 0 \\
        0 & \epsilon_{22} & 0 \\
        0 & 0 & \epsilon_{33}
    \end{pmatrix}=\begin{pmatrix}
        n_o^2 & 0 & 0 \\
        0 & n_o^2 & 0 \\
        0 & 0 & n_e^2
    \end{pmatrix}~~
\end{equation} 
where $n_o=2.21$ and $n_e=2.13$ are the ordinary and extraordinary refractive indices \cite{zelmon_infrared_1997}.

\subsection{Piezoelectric properties}

We use the stress-charge formulation to describe the piezoelectric properties of LN. This relates the stress tensor $T$, strain tensor $S$, electric field $E$, and electric displacement field $D$ according to
\begin{align}
    T&=C^E:S-e^T\cdot E\\
    D&=\epsilon^S\cdot E+e:S
\end{align}
where $C^E$ denotes the stiffness tensor at constant $E$ field, $e$ and $e^T$ denote the piezoelectric tensor and its transpose, respectively, and $\epsilon^S$ denotes the absolute RF/microwave permittivity at constant strain. We use the following values for $C^E$ $(\textrm{GPa})$, $e$ $(\mathrm{C}/\mathrm{m}^2)$, and $\epsilon^S/\epsilon_0$ (unitless) \cite{weis_lithium_1985, rodrigues_cross-polarized_2025}

\begin{align}
    C^E&=\begin{pmatrix}
        c_{11} & c_{12} & c_{13} & c_{14} & 0 & 0 \\
        c_{12} & c_{11} & c_{13} & -c_{14} & 0 & 0 \\
        c_{12} & c_{13} & c_{33} & 0 & 0 & 0 \\
        c_{14} & -c_{14} & 0 & c_{44} & 0 & 0\\
        0 & 0 & 0 & 0 & c_{44} & c_{14} \\
        0 & 0 & 0 & 0 & c_{14} & \frac{1}{2}(c_{11}-c_{12})
    \end{pmatrix}\\&=\begin{pmatrix}
        198.83 & 54.64 & 68.23 & 7.83 & 0 & 0 \\
        54.64 & 198.83 & 68.23 & -7.83 & 0 & 0 \\
        68.23 & 68.23 & 235.71 & 0 & 0 & 0 \\
        7.83 & -7.83 & 0 & 59.86 & 0 & 0 \\
        0 & 0 & 0 & 0 & 59.86 & 7.83 \\
        0 & 0 & 0 & 0 & 7.83 & 72.095
    \end{pmatrix}
    \\ \notag \\
    e&=\begin{pmatrix}
        0 & 0 & 0 & 0 & e_{15} & -e_{22} \\
        -e^{22} & e_{22} & 0 & e_{15} & 0 & 0 \\
        e_{31} & e_{31} & e_{33} & 0 & 0 & 0
    \end{pmatrix}\\&=\begin{pmatrix}
        0 & 0 & 0 & 0 & 3.65 & -2.39 \\
        -2.39 & 2.39 & 0 & 3.65 & 0 & 0 \\
        0.31 & 0.31 & 1.72 & 0 & 0 & 0
    \end{pmatrix}
    \\ \notag \\
    \epsilon^S/\epsilon_0&=\begin{pmatrix}
        \epsilon^S_{o} & 0 & 0 \\
        0 & \epsilon^S_{o} & 0 \\
        0 & 0 & \epsilon^S_{e}
    \end{pmatrix}=\begin{pmatrix}
        44 & 0 & 0 \\
        0 & 44 & 0 \\
        0 & 0 & 29
    \end{pmatrix}
\end{align}

\subsection{Electro-optic and photoelastic tensors}

The electro-optic $(r)$ and photoelastic $(p)$ tensors describe the linear change in inverse permittivity induced by an electric field and strain field, respectively. In accordance with the central theme of this paper, however, the couplings between electricity, strain, and optics are not always trivial. Therefore, we must proceed carefully to avoid double counting any couplings. For the electro-optic effect, we should utilize the clamped (constant strain) values of the electro-optic tensor $r^S$ so that the photoelastic effect does not contribute. The clamped values can be experimentally obtained through high frequency measurements, and we use the following values in $\mathrm{pm/V}$ \cite{weis_lithium_1985}:
\begin{equation}
    r=\begin{pmatrix}
        0 & -r_{22} & r_{13} \\
        0 & r_{22} & r_{13} \\
        0 & 0 & r_{33} \\
        0 & r_{42} & 0 \\
        r_{42} & 0 & 0 \\
        -r_{22} & 0 & 0
    \end{pmatrix}=\begin{pmatrix}
        0 & -3.4 & 8.6 \\
        0 & 3.4 & 8.6 \\
        0 & 0 & 30.8 \\
        0 & 28 & 0 \\
        28 & 0 & 0 \\
        -3.4 & 0 & 0
    \end{pmatrix}~~.
\end{equation}

Similarly, we must use the photoelastic coefficients under constant electric field $(p^E)$ if we include the electro-optic effect separately. Unlike for the electro-optic effect, however, these values cannot be measured directly. Firstly, photoelastic coefficients including an electro-optic contribution can be measured experimentally. For our mechanical only simulation, we use the values given in \cite{rodrigues_stimulated_2023, rodrigues_cross-polarized_2025}, which include the secondary electro-optic effect:
\begin{align}
    p&=\begin{pmatrix}
        p_{11} & p_{12} & p_{13} & p_{14} & 0 & 0 \\
        p_{12} & p_{11} & p_{13} & -p_{14} & 0 & 0 \\
        p_{31} & p_{31} & p_{33} & 0 & 0 & 0 \\
        p_{41} & -p_{41} & 0 & p_{44} & 0 & 0 \\
        0 & 0 & 0 & 0 & p_{44} & p_{41} \\
        0 & 0 & 0 & 0 & p_{14} & \frac{1}{2}(p_{11}-p_{12})
    \end{pmatrix}\\&=\begin{pmatrix}
        -0.026 & 0.088 & 0.134 & -0.083 & 0 & 0 \\
        0.088 & -0.026 & 0.134 & 0.083 & 0 & 0 \\
        0.177 & 0.177 & 0.07 & 0 & 0 & 0 \\
        -0.151 & 0.151 & 0 & 0.145 & 0 & 0 \\
        0 & 0 & 0 & 0 & 0.145 & -0.151 \\
        0 & 0 & 0 & 0 & -0.083 & -0.057
    \end{pmatrix} ~~.
\end{align}

To explicitly include the electro-optic effect in our simulation, we must then remove the electro-optic contribution from the photoelastic tensor. This contribution strongly influences the measurements of $p_{13}$ and $p_{33}$, so we change them according to $p_{13}^E=p_{13}-0.043$ and $p_{33}^E=p_{33}-0.154$ \cite{weis_lithium_1985}.

\end{bibunit}

\end{document}